\documentclass[epj]{webofc}
\usepackage[varg]{txfonts}   % Web of Conferences font
\usepackage{graphicx,color} % for figures
\usepackage{bm}       % bold math 
\usepackage{amsmath}  
\usepackage{amssymb}
\woctitle{21st International Conference on Few-Body Problems in Physics}
\begin{document}
\title{Mesonic Decay of Charm Hypernuclei $\Lambda^+_c$}
\author{Sabyasachi Ghosh\inst{1}\fnsep\thanks{\email{sabyaphy@ift.unesp.br}},
        Carlos E. Fontoura\inst{1}\fnsep\thanks{\email{eduardo@ift.unesp.br}}, 
        Gast\~ao Krein\inst{1}\fnsep\thanks{\email{gkrein@ift.unesp.br}},
        }
        
\institute{Instituto de F\'{\i}sica Te\'orica, Universidade Estadual Paulista \\
Rua Dr. Bento Teobaldo Ferraz, 271 - Bloco II, 01140-070 S\~ao Paulo, SP, Brazil 
          }

\abstract{$\Lambda^+_c$ hypernuclei are expected to have binding energies and 
other properties similar to those of strange hypernuclei in view of the similarity between 
the quark structures of the strange and charmed hyperons, namely $\Lambda(uds)$ and 
$\Lambda^+_c (udc)$. One striking difference however occurs in their mesonic decays, as there is 
almost no Pauli blocking in the nucleonic decay of a charm hypernucleus because the final-state
nucleons leave the nucleus at high energies. The nuclear medium nevertheless affects the 
mesonic decays of charm hypernucleus because the nuclear mean fields modify the masses of 
the charm hyperon. In the present communication we present results of a first investigation 
of the effects of finite baryon density on different weak mesonic decay channels of the 
$\Lambda^+_c$ baryon. We found a non-negligible reduction of the decay widths as compared 
to their vacuum values. 
}
\maketitle
\section{Introduction}
\label{intro}

The possible existence of charm hypernuclei has been suggested~\cite{Tyapkin} right 
after the discovery of the charm quark and a first explicit  calculation~\cite{DoverKahana} 
of their binding energies was performed using a meson exchange model with coupling constants 
determined by SU(4) flavor symmetry. More recently, calculations were 
performed~\cite{TsushimaKhanna} using the quark meson coupling model at the Hartree 
level {\textemdash} the predicted binding energies per baryon in $\Lambda^+_c$, $\Sigma_c$, 
$\Xi_c$ hypernuclei ranging from O to Pb are quite substantial, on average of the order of 
$7.5$~MeV. Such binding energies are of similar magnitude to those of $J/\Psi$ binding in
nuclear matter~\cite{Krein:2010vp} and finite nuclei~\cite{Tsushima:2011kh}. Vacuum
effects~\cite{Krein:1993jb} and Fock terms~\cite{Krein:1998vc}, although crucial for 
capturing pionic effects in matter~\cite{Bracco:1998qw}, are not expected to have a great 
impact on the binding energies. Up to now there is no experimental evidence of the existence 
of a charm hypernucleus, apart from an emulsion experiment at Fermilab with $250$~GeV protons 
that indicated that $\Lambda^+_c$ hypernuclei have been produced in the collisions~\cite{exp}. 
New experimental possibilities were explored~\cite{Bressani,Buyatov} at the beginnings of 
the 1990's, but concrete hopes for the production of such hypernuclei comes with the 
completion of the FAIR facility (where $\Lambda^\pm_c$ can be produced in $\bar pp$ 
annihilation processes~\cite{Haidenbauer:2009ad}) and the installation of a $50$~GeV 
high-intensity proton beam at the J-PARC complex. 

The study of charm hypernuclei is interesting for several reasons; among the most
interesting in our view are:  (1) they bring the opportunity to learn about the poorly 
known interactions of charmed baryons with nucleons~\cite{{Haidenbauer:2007jq},Fontoura:2012mz}; 
(2) they offer also the opportunity to learn about medium effects on diquark 
correlations, as the masses of the light $u$ and $d$ quarks are affected by the 
medium while the mass of the heavy $c$ quark is not affected; and (3) few-nucleon 
(typically one or two) charmed states have the potential of having a rich 
spectroscopy similar to the X,Y,Z mesons. 

One particular observable that can be affected by the medium is the weak-decay lifetime 
of the charmed baryon because of the change in its mass. There are no Pauli blocking 
effects on the decaying baryon -- for the case of $\Lambda^+_c$, for instance, the decays 
$\Lambda\pi^+$ and $\Sigma^+\pi^0$ are not blocked, and in the decay $p{\bar K}^0$ the proton 
leaves the nucleus with a large momentum. In the present communication we present the results 
of a first investigation~\cite{us} of the effects of finite baryon density on different weak 
mesonic decay channels of the $\Lambda^+_c$ baryon.

\section{In-medium mesonic decays of $\Lambda^+_c$}
\label{decay}

\begin{figure}[h!]
\centering
\includegraphics[scale=0.5]{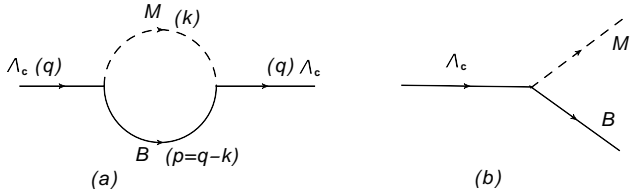}
\caption{(a) One loop baryon-meson $BM$ self-energy of $\Lambda^+_c$: $B = \Lambda, \Sigma^+, p$;
$M = \pi^+, \pi^0, {\bar K}^0$. (b) decay diagram, the imaginary part of (a).} 
\label{fig:1}
\end{figure}

We evaluate the decay rate in infinite nuclear matter via the imaginary part of the
self-energy diagram depicted in Fig.~\ref{fig:1}. To evaluate the diagram, we employ the
effective weak-decay Lagrangian density
\begin{equation}
{\cal L}_{\Lambda_c BM} = ig_{\Lambda_c BM}
\bar{\psi}_{\Lambda_c}\left(A_{BM}+B_{BM}\gamma^5\right)
\phi_M\psi_B + {\rm h.c.},
\label{lag_W}
\end{equation}
where $g_{\Lambda_c BM}= 10^{-2} G_FV_{ud}V_{cs}$, with $G_F=1.16\times 10^{-5}$ GeV$^{-2}$ 
being the Fermi constant and $V_{ud}=0.974$, $V_{cs}=0.973$ relevant CKM matrix elements.
For the density dependence of $\Lambda^+_c$ we use the results from the QMC 
model~\cite{TsushimaKhanna}. Finite nuclei can be treated in the local density approximation.

\vspace{0.5cm}

We start examining the role of the Pauli principle in the $\Lambda^+_c\rightarrow p{\bar K}^0$ decay,
neglecting all other possible medium effects.  The in-medium decay width $\Gamma_{T}(q)$ for a 
generic baryon-meson (BM) decay $\Lambda^+_c \rightarrow BM$, where $q = |{\bf q}|$ with ${\bf q}$ 
the momentum of the $\Lambda^+_c$, can be written as the difference $\Gamma_T(q) = \Gamma_V(q) - \Gamma_D(q)$, where $\Gamma_V(q)$ is vacuum decay width and $\Gamma_D(q)$ is a density 
dependent width due to Pauli-blocking:  
\begin{eqnarray}
\left(\begin{array}{l}
\Gamma_V(q) \\
\Gamma_D(q) \end{array} 
\right)
&=&  \frac{g^2_{\Lambda_c BM}}{16\pi q m_{\Lambda_c}}
\left[ \left(m_{\Lambda_c} + m_B\right)^2 - m^2_M \right] 
\left[ |A_{BM}|^2 + \left( \frac{\left(m_{\Lambda_c}-m_B\right)^2 - m_M^2}
{\left(m_{\Lambda_c} + m_B\right)^2-m^2_M} \right)\, |B_{BM}|^2 \right] \, 
\nonumber \\
&& \times \, \int^{\omega^+(q)}_{\omega^-(q)} 
\left(\begin{array}{c}
1 \\
\theta(\omega - \omega^{\rm th}(q))\end{array} 
\right) \, d\omega,
\label{Gammas}
\end{eqnarray}
where $m_{\Lambda_c}$, $m_B$ and $m_M$ are the masses of $\Lambda^+_c$, baryon $B$ 
and meson $M$, respectively, and  $\theta$ is the step function with  
\begin{equation}
\omega^{\pm}(q) = \frac{R^2}{2m_{\Lambda_c}^2}\left[(q^2 + m_{\Lambda_c}^2)^{1/2}
\pm q W \right],\hspace{1.0cm}
\omega^{\rm th}(q) = \left(q^2 + m^2_{\Lambda_c}\right)^{1/2} - 
\left(k^2_F + m^2_B\right)^{1/2},
\label{omega-th} 
\end{equation}
where $R^2=m_{\Lambda_c}^2+m_M^2-m_B^2$, $W = (1-4m^2_M m^2_{\Lambda_c}/R^4)^{1/2}$,
and $k_F$ is the nuclear matter Fermi momentum given in terms of the baryon density 
$\rho$ as $k_F = (3\pi^2/2 \, \rho)^{1/3}$.

\begin{figure}[h!]
\begin{center}
\includegraphics[scale=0.3]{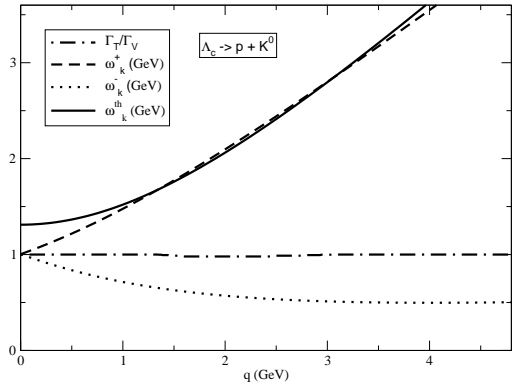}
\caption{Momentum dependence of the ratio of the total to vacuum decay widths 
(dash-dotted line) for the $\Lambda_c\rightarrow p{\bar K}^0$ decay channel
in nuclear matter at the saturation density ($\rho = \rho_0 = 0.16~{\rm fm}^{-3}$).  
Also shown are the quantities $\omega^{+}(q)$ (dashed line), $\omega^{-}(q)$ 
(dotted line), and $\omega^{\rm th}(q)$ (solid line). }
\label{gm_k_Lc2}
\end{center}
\end{figure}

There is no Pauli blocking in the $\Lambda^+_c \rightarrow \Lambda \pi^+$ 
and $\Lambda^+_c \rightarrow \Sigma^+\pi^0$ decays and $\Gamma_D(q)$ is nonzero only 
in the nucleonic decay $\Lambda^+_c \rightarrow p{\bar K}^0$. Notice that $\Gamma_D(q)$ 
vanishes for $\rho_B = 0$ as $\omega^{\rm th}(q) > \omega^{+}(q)$ in this case and the 
step function in Eq.~(\ref{Gammas}) gives zero for the integral. The dot-dashed line in Fig.~\ref{gm_k_Lc2} 
presents the momentum dependence of the ratio $\Gamma_T/\Gamma_V = 1 - \Gamma_D/\Gamma_V$ 
for the $\Lambda^+_c \rightarrow p{\bar K}^0$ decay channel in nuclear matter 
($\rho = \rho_0 = 0.16~{\rm fm}^{-3}$). Clearly, the effect of the Pauli
principle is negligible. Physically, this is due to the fact that the momentum of the
outgoing proton is of the order of, or larger than the Fermi momentum {\textemdash} this is
transparent in the plot of $\omega^{\rm th}(q)$ that shows that it barely takes values
within the region delimited by the curves of $\omega^{+}(q)$ and $\omega^{-}(q)$ 
as demanded by the theta function in Eq.~(\ref{Gammas}) for a $\Gamma_D(q) \neq 0$.

\begin{figure}[t!]
\begin{center}
\includegraphics[scale=0.3]{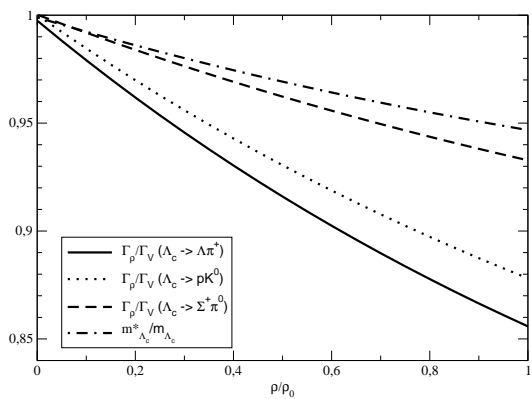}
\caption{$\Gamma_\rho/\Gamma_V$ for $\Lambda_c\rightarrow\Lambda\pi^+$ (solid line),
$\Lambda_c\rightarrow\Sigma^+\pi^0$ (dashed line), $\Lambda_c\rightarrow p{\bar K}^0$ 
(dotted line) decay channels with $\rho/\rho_0$. Also shown is 
$m^*_{\Lambda_c}/m_{\Lambda_c}$ (dot-dashed line). }
\label{gm_dense}
\end{center}
\end{figure}

Next, we consider the effect of the in-medium mass shift of $\Lambda^+_c$ on
the decay widths of the different channels {\textemdash} as the proton leaves the 
nucleus with high momentum, its mass shift is negligible. The density dependence
on the mass is taken from the QMC model~\cite{TsushimaKhanna}, which can be 
parametrized as 
\begin{equation}
m^*_{\Lambda_c}(\rho)/m_{\Lambda_c} = a_0 \, e^{-a_1\, \rho/\rho_0} + a_2,
\label{m*}
\end{equation}
with $a_0=0.121$, $a_1=0.565$ and $a_2=0.878$. The in-medium decay width is denoted by
$\Gamma_\rho$ and is given by the same expression as $\Gamma_V$ in Eq.~(\ref{Gammas})
but with $m_{\Lambda_c}$ replaced by $m^*_{\Lambda_c}$. In Fig.~\ref{gm_dense} we present
the ratio $\Gamma_\rho/\Gamma_V$ as function of the baryon density, where 
$\Gamma_\rho$ is evaluated at $q=m^*_{\Lambda_c}$ and $\Gamma_V$ is evaluated 
at $q=m_{\Lambda_c}$. At the nuclear matter saturation density, $m^*_{\Lambda_c}=2.146$~GeV. 
The in-medium reduction of the decay widths is not small. Specifically, at the normal 
nuclear matter density, the reductions are $14.5\%$, $12\%$ and $6.5\%$ for $\Lambda_c \rightarrow \Lambda\pi^+, \Sigma^+\pi^0, \,{\rm and} \; p{\bar K}^0$ decay channels, respectively. 
A local density approximation leads to similar reductions for medium to heavy nuclei~\cite{us}. 

Several issues need careful examination before definite conclusions can be drawn. 
Specifically, SU(4) flavor symmetry breaking in meson-baryon couplings have been 
shown to be significant~\cite{ElBennich:2011py} and its impact on the results needs 
to be assessed {\textemdash} the subject is also of crucial importance for the $DN$ 
interaction~\cite{Fontoura:2012mz} in connection to $D$-mesic nuclei. The consequences 
of medium modifications on the mass of $\Lambda^+_c$ on non-mesonic weak decays of 
charmed hypernuclei is under investigation~\cite{us2}.

\begin{acknowledgement}
Work partially by Funda\c{c}\~ao de Amparo \`a Pesquisa do Estado de S\~ao Paulo-FAPESP,  
Grants 2012/16766-0 (S.G.) and No. 2013/01907-0 (G.K.), and Conselho Nacional de Desenvolvimento 
Cient\'{\i}fico e Tecnol\'ogico - CNPq, Grant No.305894/2009-9 (G.K.). The work of C.E.F. was 
supported by a scholarship from Universidade Estadual Paulista.
\end{acknowledgement}

%
% BibTeX or Biber users please use (the style is already called in the class, ensure that the "woc.bst" style is in your local directory)
% \bibliography{name or your bibliography database}
%
% Non-BibTeX users please use
%

\end{document}